\DocumentMetadata{pdfversion=1.2}
\nonstopmode\synctex=1
\documentclass{IEEEtran}

\usepackage{tikz, bbm, xcolor, amsmath, amssymb, listings}
\usepackage[ruled, vlined, linesnumbered, longend]{algorithm2e}
\usetikzlibrary{intersections, decorations.pathreplacing}
\tikzset{ >= latex, }
\usepackage[inline]{enumitem}

\newcommand{\tht}{\theta}
\newcommand{\del}{\partial}
\newcommand{\R}{\mathbb R}

\usepackage{soul}

\title{Neural directional distance field object representation for
uni-directional path-traced rendering}
\author{Anonymous}
\author{Annada Prasad Behera\textsuperscript{1,2}
\qquad Subhankar Mishra\textsuperscript{1,2}\vspace{.1in}\\
{\small
\textsuperscript{1}National Institute of Science Education and Research, Bhubaneswar\\
\textsuperscript{2}Homi Bhabha National Institute, Mumbai\\
}}

\begin{document}
\maketitle

\begin{abstract}Faster rendering of synthetic images is a core
problem in the field of computer graphics. Rendering algorithms, such as
path-tracing is dependent on parameters like size of the image, number
of light bounces, number of samples per pixel, all of which, are fixed if
one wants to obtain a image of a desired quality. It is also dependent on
the size and complexity of the scene being rendered. One of the largest
bottleneck in rendering, particularly when the scene is very large,
is querying for objects in the path of a given ray in the scene. By
changing the data type that represents the objects in the scene, one
may reduce render time, however, a different representation of a scene
requires the modification of the rendering algorithm.  In this paper,
(a) we introduce directed distance field, as a functional representation
of a object; (b) how the directed distance functions, when stored as a
neural network, be optimized and; (c) how such an object can be rendered
with a  modified path-tracing algorithm.\end{abstract}

\section{Introduction} Path tracing is used for faithfully reproducing
photo-realistic images from a mathematical description of a 3D scene.
However, path tracing algorithm has very high resource requirements and
it takes a long time to render a image. As the demand for real time
rendering increases, the field has seen a lot of interest in recent
years, and many techniques have been proposed for making the rendering
faster, by adding hardware resources like more powerful GPUs, vectorizing
computation on hardware\cite{zhou2021vectorization} and using learning
algorithms like the deep learning supersampling\cite{xiao2020neural}.

Nevertheless, the core path tracing algorithm has seen little change
towards faster rendering. One bottleneck in path tracing algorithm
occurs when the algorithm computes ray-triangle intersection against
every triangle in the scene. A linear search for the nearest triangle
in the scene becomes extremely slow. Use of data structures like bounding
volume hierarchies (BVH), Octrees and Kd-trees \cite{karras2012maximizing}
has helped in reducing it to log-linear time. In this paper, we propose
a data structure that we call the \emph{directed distance function} that
theoretically takes a constant time to compute the nearest ray-triangle
intersection, albeit with a preprocessing time for building the data
structure.

The directed distance field can be used to represent individual objects
as well as entire scene. Directed distance fields are dense fields on
five-dimensional space and hence representing them is also not trivial
since a discretization will consume $O(n^5)$ space. The field can be
stored as a pure function, which although extremely fast, it hard to build
(for artists). In this paper, we explore neural networks as a universal
function estimator to represent the field.

To estimate the field with neural networks, we need training data. The
field around the object is complex which necessitates a systematic
sampling technique for building the dataset to train on. Otherwise the
network fails to learn the distance field of the object. We introduce
sampling technique to build the dataset and observe that it performs
better than sampling techniques used in similar works.

Rendering algorithms are generally tied to the underlying scene
representations. For example, sphere tracing for signed distance fields
and ray-triangle intersections for polyhedral mesh.  Hence, we propose a
modification to the path tracing algorithm to render a directed distance
field. We modify the path tracing algorithm to use this data structure.

In this paper, our contribution are as follows.
\begin{enumerate}[label=(\alph*)\;]
\item We analyze directed distance function as a alternative representation
of object for path-trace rendering.
\item We provide modification of the path-tracing rendering algorithm
to be able to render DDFs, which are represented as neural networks.
\item We provide a informative sampling techniques to build datasets to
optimize the above given neural network.
\end{enumerate}

\section{Related Works} The rendering algorithm depends largely on the
the underlying data structure that represents the scene. Therefore,
we discuss different representations in this section.

\textbf{Explicit representation.}\;In explicit scene representation,
the geometry is defined in terms of the explicit location of the
objects and its properties in the scene. Such explicit representation
include point clouds, meshes, voxels which include the position of the
vertices and their color. The rendering algorithm makes direct queries
about these properties in the scene, while rendering. Since the explicit
representation is compact, easy to represent and reason with while also
providing easier ways to model the scene (for artists.) However, their
topology is not easily handled, for example, performing boolean operation
like union, intersection in constructive solid geometry, or in computer
vision, for learning algorithms that reconstruct geometry from images
becomes hard and, sometimes impossible.

\textbf{Implicit representation.}\ The geometry can be defined without
explicitly defining the location and color of objects in the scene,
and using the functions instead. The most popular of such implicit
representation is the signed distance function.  Signed distance functions
(sometimes, just distance field or SDFs), is a field over the 3D space
$\R^3$, where every point in the field is defined as the distance to
the nearest surface.  The sphere tracing algorithm\cite{hart1996sphere}
is used to render SDF. However, storing SDFs as discretized values
comes with a huge space cost and hence have seen very limited use or are
manually crafted by artists with mathematical functions. It's resurgence
in recent years is due to learning algorithms that can represent arbitrary
functions and can be learned. The zero level set of a signed distance
function represents the surface.

Other implicit function representations that are popular in learning
algorithms are radiance fields\cite{2021nerf}, that are like SDF but instead
of distances, the field values stores the color and density (or
opacity) of the point. They are particularly suited for volumetric
rendering. If, instead the points only represent whether it is
inside or outside a surface, the field is occupancy function.  These
functions are learned with universal function approximators like neural
networks\cite{hornik1989multilayer} and can be converted into explicit
representation by marching cubes\cite{chernyaev1995marching} algorithm.

\textbf{Directional distance function.}\ Signed distance functions with
directional component has been used in prior works\cite{aumentado22,
yenamandra22, liu17, zins22, zobeidi21} for shape representation and
reconstruction using learning algorithms. Truncated Signed Directional
Distance function\cite{liu17} have been used for pose estimation. The
geometric properties of DDFs, and differentiable rendering of
DDFs by using a probabilistic directional distance field with $C^1$
consistency for are discussed in \cite{aumentado22}. However, they are
ray-traced rendered instead of path-traced. Although this technique
is suitable for reconstruction, it cannot be used for multiple bounce
path-traced physically based rendering. Directional distance function
are used for representation and reconstruction of shapes that are
to be optimized for rendering with learning algorithms like neural
networks with single images\cite{aumentado22, zins22}, with ground
truth meshes\cite{aumentado22, yenamandra22} and with multiple images
and their depth fields\cite{zins22}.

\section{Background} In this section, we introduce the directed distance
function, it's mathematical properties and the rendering pipeline, as
required in the later sections.

\subsection{Directional signed distance functions} Let $S$ be the surface
of a 3D object which lies entirely within the bounded box $B\subset\R^3$.
A directional distance field (DDF) $\phi:B\times\R^2\to\R^+$ takes
point, $x\in B$ and direction, $\tht\in\R^2$ (henceforth \emph{viewing
direction}) and returns the minimum distance $d$, from the point $x$,
to the surface $S$, in the direction $\tht$.  A binary visibility
function $\xi(x, \tht) = \mathbbm 1_S(x,\tht)$ returns 1 if the ray $x$ in
direction $\tht$ intersects the surface $S$. The oriented point $(x,\tht)$
is said to be \emph{visible} if $\mathbbm 1_S(x,\tht)$ returns 1. In
the following paragraphs, we present some geometric properties of DDF
(proof of which maybe found at \cite{aumentado22, zobeidi21}).

For a given point, $x$, any viewing angle, $\tht$ can be represented in
two degree of freedom, with \emph{azimuthal} $\tht_0$, and \emph{polar}
$ \tht_1$ angles, for example. When specified with those angles, moving
$t$ distance from $x$ in $\tht$ direction is, \begin{equation}
	x' = x+t\begin{bmatrix}
		\cos(\tht_0)\sin(\tht_1)& \sin(\tht_0)\sin(\tht_1)&
		\cos(\tht_1)\end{bmatrix}^T
\end{equation} We abuse the notation and write $x+t\tht$ to refer to
the point at $t$ distance from $x$ in the viewing direction, $\tht$,
as this way is popularly used in ray marching computations. The actual
computation is implementation-specific.

\textbf{Directed Eikonal Equation.} For any visible oriented point
$(p,\tht)$, the following holds,
\begin{equation}
	\del_x\phi(x,\tht)\cdot\tht  = -1
\end{equation}
In the viewing direction, for every marching step the DDF value decreases
by the step size, i.e,  the DDF, $\phi(x, \tht)-\phi(x+t\tht, \tht) =
t$, which follows that the maximum rate of change in $\phi$ occurs in
the direction opposite to the viewing direction.

\textbf{Surface normal.} The point on the surface $S$ is given as,
$q(x,\tht) = \phi(x,\tht)\tht+x \in S$.  The normal at the given surface
is given by,
\begin{equation}\label{eq_ddfnormal}
	n(x,\tht) = \frac{\kappa\del_x\phi(x,\tht)^T}{\lVert\del_x\phi(x,\tht)\rVert_2}
\end{equation}
where $\kappa$ is chosen from $\{-1,1\}$ such that $n^T\tht<0$.

\textbf{Gradient consistency.} From any visible oriented point $(x,\tht)$,
a infinitesimal change in the viewing direction, $\delta\tht$ effectively
pushes the position $x$ in the direction, $\tht$, i.e,
\begin{equation}
	\del_\tht\phi(x,\tht) = \phi(x,\tht)\del_x\phi(x,\tht)\delta\tht
\end{equation}
where $\del\tht=\omega\times\tht$. This relates the gradient of $\phi$
with a rotational perturbation $\del\tht$ with respect to both position
and direction.

\textbf{Unsigned distance field.} The distance field (like signed distance
function, except that the field is not negative inside the surface),
can be recovered from the DDF by computing the minimum of DDF against
all the direction for any given point $x$,
\begin{equation}
\label{eq_udf}\text{UDF}(x) = \min_{\tht} \phi(x, \tht)
\end{equation}
The UDF (unlike DDF) has no discontinuities.

\textbf{Locally differentiable} For any visible oriented point $(x,\tht)$,
the geometry of 2D manifold $S$ near $q$ is completely characterized
by $\phi(x,\tht)$ and it's derivatives. This allows us to recover the
surface properties like normals and curvature from the DDF only.

The above mentioned properties will be used to modify the path
tracing algorithm to use DDF to render images. The reader may refer to
\cite{aumentado22, zobeidi21} for a more general and detailed treatment
of DDF.

\section{Methodology}
\begin{figure}
\centering
\begin{tikzpicture}[>=latex]
\usetikzlibrary{shapes.misc}
\path	(0.5, 1.8)
		node (co)[coordinate]{}
		node [draw, rectangle, inner sep=3pt, rotate=37.5]{}
		node[above=3pt]{camera, $c_o$}
	(3, 1.8)
		node [cross out, draw]{}
		node [cross out, draw, rotate=45]{}
		node [fill=white, draw=black, circle, inner sep=1pt]{}
		node (em)[inner sep=5pt, circle]{}
		node [above=5pt]{lamp, $e$}
	(2, 0)
		node (h0)[coordinate]{}
		node [below]{$x$}
	(5.5, 1.4)
		node (h1)[coordinate]{}
		node [right]{$x'$}
;
\fill [gray!40]
	(h0) ++(.75, 0) arc(0:180:.75)
	node [black, pos=.9, right]{$\Omega$}
	node [black, pos=.25, right]{$\omega_i$}
	;
\fill [gray!40]
	(h1) ++(0, .75) arc(90:270:.75)
	node [black, pos=.1, below] {$\Omega'$};

\foreach \i in {.05, .125, ..., .35} {
	\path   (h0) ++(.75, 0) arc(0:180:.75)
		node(w) [pos=\i, coordinate]{};
	\fill (w) circle (1pt);
}

\path (co) -- node(cd)[pos=.04, coordinate]{} (h0);
\draw [->, thick]
	(h0) -- node[left]{$L_o$}
	(cd) node[fill, circle, inner sep=1pt]{};
\draw [->, gray, thick](em) -- node[left, pos=.3, black]{$L_e$}(h0);
\draw [->, thick](em) -- node[above]{$L_e'$}(h1);
\draw [<-, thick](h0) -- node[below]{$L_i$}(h1);
\draw (0, 0) -| (5.5, 2.2);
\end{tikzpicture}
\caption{The 2D cross-section of a scene: With respect to the rendering
equation demonstrates the path the light takes at multiple after bounces
in the and the radiance computation over the hemisphere, $\Omega$.}
\label{fig_geo}
\end{figure}
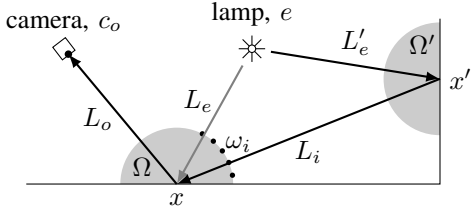

\subsection{Object representation}
\textbf{Neural network.} When the
DDF is stored as a neural network, the DDF is learned either from a
set of multiple images\cite{yoshitake23} or ground truth scene. We will
only discuss ground truth mesh in this work. Consider a neural network,
$\text{MLP}_w:(\R^3,\R^2)\to\R$ that takes a oriented point $(x,\tht)$ and
returns the DDF field value. Given a triangle mesh scene, the training
dataset consists of sampling points from bounding box. After the weights
$w$ are optimized, the $\text{MLP}_w$ can be used for rendering the scene.

For invisible oriented points, the field value $\phi(x,\tht)$ is set
to $\infty$, however this causes problem in optimizing the network for
the object. Therefore a squishing function maybe used for learning the
field value, with the loss function,
\begin{equation}
	L_w = \sigma(y) - \sigma(\tilde y)
\end{equation}
{where, $y=\phi_\text{gt}(x,\tht)$ is the ground truth DDF value, $\tilde
y=\text{MLP}_w(x,\tht)$ is the network representation of the object} and
the squish function\cite{zobeidi21} $\sigma(x)$ is a strictly increasing
function that approaches a finite value as input $x$, approaches $\infty$,
like tanh$(x)$ or erf$(x)$.

\subsection{Surface normal computation} The normal is required at each
intersection of the ray with the scene for shading. The normal of a
directed distance field can be computed using Eq. \ref{eq_ddfnormal} by
exploiting the fact that the distance function is locally differentiable.
As the DDF is stored as neural network, the normals are trivial to compute
with automatic differentiation,
\begin{equation}
    \del_x\phi(x,\tht) = \del_x\text{MLP}_w(x,\tht)
\end{equation}
{ taking care to find the derivative with respect to the input point $x$,
rather than the weights $w$ as done in general backward pass. In PyTorch,
the relevant code will be,}
\begin{lstlisting}
    ddf = MLP(position, direction)
    ddf.backward()
    normal = position.grad
\end{lstlisting}
{ The $\kappa$ is assigned to make the normal point outward, using the dot product,
between the normal and viewing direction such that, $n^T\tht<0$.}

\section{Dataset, network architecture, and renderer}
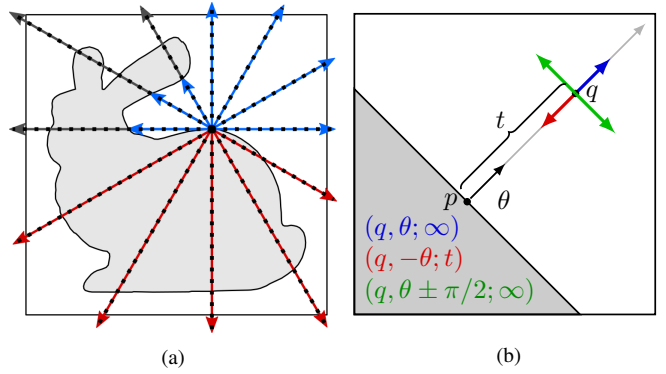
\begin{figure}
\begin{minipage}{.49\linewidth}
\centering
\includegraphics[width=\textwidth]{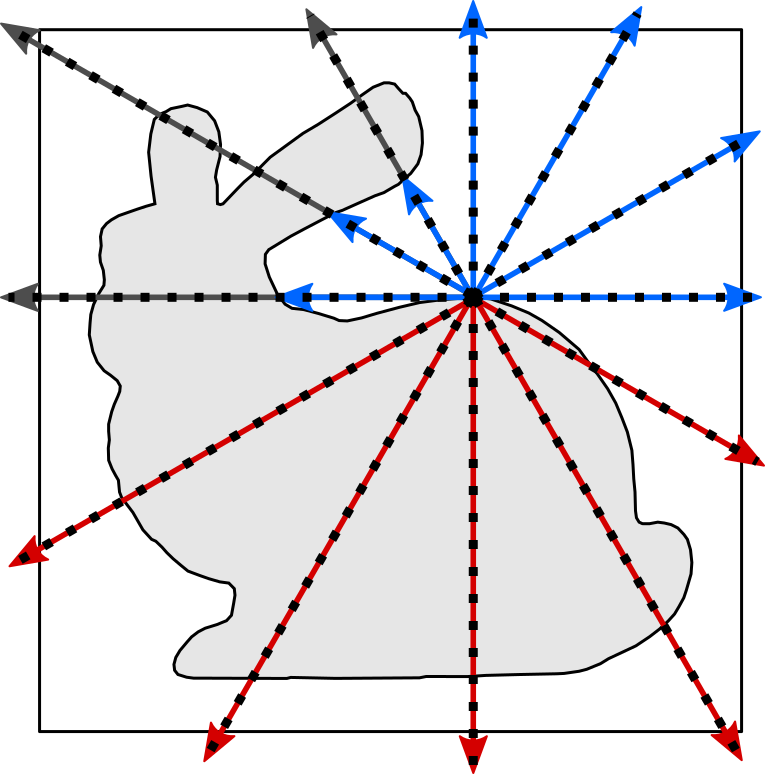}\\{\footnotesize (a)}
\end{minipage}
\begin{minipage}{.49\linewidth}\centering
\tikz[line width=.65pt, >=latex]{
    \fill[gray!40, draw=black] (0, 3) --
        node[pos=.5, coordinate](p){}
        node[pos=.5, fill=black, circle, inner sep=1pt](p){}
        node[pos=.5, left, black]{$p$}
        (3, 0) -- (0,0) -- cycle;
    \draw (0,0) rectangle (4,4);
    \draw[white] (0, -.2) -- (0, -0.2);
    \draw [->, black!30] (4,4) node(d){} (p) --
        node[pos=.2,coordinate](n){}
        node[pos=.4,coordinate](mb){}
        node[pos=.6,coordinate](m){}
        node[pos=.8,coordinate](ma){} (d);
    \draw [->] (p) -- node[below right]{$\tht$} (n) ;
    \draw [->, red!85!black, very thick] (m) node[right, black]{$q$} -- (mb);
    \draw [->, blue!85!black, very thick] (m)
        node[fill=black, circle, inner sep=1pt]{} -- (ma);
    \draw [->, green!75!black, very thick, rotate around={135:(m)}] (m) -- ++(.75, 0);
    \draw [->, green!75!black, very thick, rotate around={-45:(m)}] (m) -- ++(.75, 0);
    \draw [decorate, decoration={brace, raise=5pt}]
        (p)-- node[above left=3pt]{$t$}(m);
    \node[text width=1in, above right](0,0){
        $\color{blue!85!black}(q, \tht; \infty)$
        $\color{red!85!black}(q, -\tht; t)$
        $\color{green!55!black}(q, \tht\pm\pi/2; \infty)$
    };
}\\{\footnotesize(b)}
\end{minipage}
\caption{(a) Sampling strategy: the object, bunny rabbit is shown in gray
in the bounding box. The points sampled from the ray shown in red and
gray are rejected, while the points on ray in blue is accepted into the
training data. (b) collecting data-points on rays from the surface: at point,
$q = p+\tht{}t$, the distance to the surface in the direction $-\tht$ is $t$,
and in the direction $\tht$, { if it doesn't hit any other face,}
the field value is infinite.}
\label{fig_sampling}
\end{figure}

\subsection{Neural network architecture and training}
The neural network is written in Pytorch. The network is a fully connected
network with 8 weight-normalized hidden layers of width 512 followed by
ReLU activation and dropout (with probability, $p=0.5$). The network
was optimized with Adam optimizer, with learning rate $10^{-7}$ over
the clamp error loss defined as,
\begin{align}
L &= \lvert\text{clamp}(y,\delta) - \text{clamp}(\tilde y,\delta) \rvert
\end{align}
where $\text{clamp}(x,\delta):=\min(\delta,\max(-\delta, x))$ {
is the squishing function} with hyperparameter $\delta$ to control the
distance from the surface over which we optimize the network over. Larger
values of $\delta$ helps in ray tracing the object from larger distances,
while the smaller values helps concentrate the network capacity on
surface details. This is a trade-off between surface details and draw
distance for ray-tracing.

For training the network, we generate the training data by sampling
points from the bounding box, $B$. Uniform random sampling the points
and viewing directions trains the network poorly because many oriented
points have infinite values and hence, do not contribute to the shape
estimation. Therefore, we have used a more informative sampling over
the surface to learn from, as described in the next section.

\subsection{Dataset preparation}
The directed distance field changes rapidly around the object it
represents, while also being in five-dimension. Therefore without a
dense sampling of the field values around the point, the neural network
cannot optimize the object shape.  We start from the ground
truth triangle mesh containing vertices and faces.  We use three
hyperparameters for sampling the training data points, face samples
$s_\text{fc}$, direction samples $s_\text{dr},$ and marching samples
$s_\text{p}$.

To sample a point $q$, from a face with vertices $\{p_0, p_1, p_2\}$,
we first sample two real values from the uniform distribution $U$, and use
the barycentric co-ordinates to get points on the face.
\begin{align}
\begin{split}
    S_\text{fc} = \{&(1-a-b)p_0 + ap_1 + bp_2:\\
    &a,b \sim U[0,1]\},\;|S_\text{fc}|=s_\text{fc}
\end{split}
\end{align}
And for sampling directions, we sample from a uniform distribution and convert it
to polar angle $\tht_0$, and azimuthal angle $\tht_1$.
\begin{align}
\begin{split}
    S_\text{dr} = \{&\tht_0 = a\pi,\;\tht_1 = 2b\pi:\\
    &a,b \sim U[0,1]\},\;|S_\text{dr}|=s_\text{dr}
\end{split}
\end{align}
For each points on the face $p\in S_\text{fc}$, and direction
$\tht\in S_\text{dr}$, we march the ray in steps of $t$ and collect the
field values into the training dataset $D$ (Figure \ref{fig_sampling}),
\begin{align}
    D \leftarrow (p+t\tht, -\tht; t)\cup D,\quad\text{for}\;t>0
\end{align}
except for two conditions: (a) if the ray marches into the mesh, where
by definition the field value is not defined (red rays) or, (b) the ray
hits another part of the mesh, (gray rays, in Fig.\ref{fig_sampling}).
In either case, we have used odd-even rule to determine whether the
direction we are sampling is inside or outside the mesh, using the fact
that when is outside the bounding box of the mesh, the ray is always
also outside the mesh. In case, the ray never enters the mesh, we also
collect the following values too,
\begin{align}
    D \leftarrow (p+t\tht, \tht; \infty)\cup D,\quad\text{for}\;t>0\\
    \label{eq_perp_sample}D \leftarrow (p+t\tht, \tht\pm \pi/2;\infty)\cup
    D,\quad\text{for}\;t>0
\end{align}
We sample according to Eq. (\ref{eq_perp_sample}) only when the angles perpendicular
to $\tht$ do not hit any face. We ignore when it hits the face because,
in that case, the rays sampled from the face it hits will cover field
value at the point.

\section{Result and evaluation}
\begin{figure*}[h]
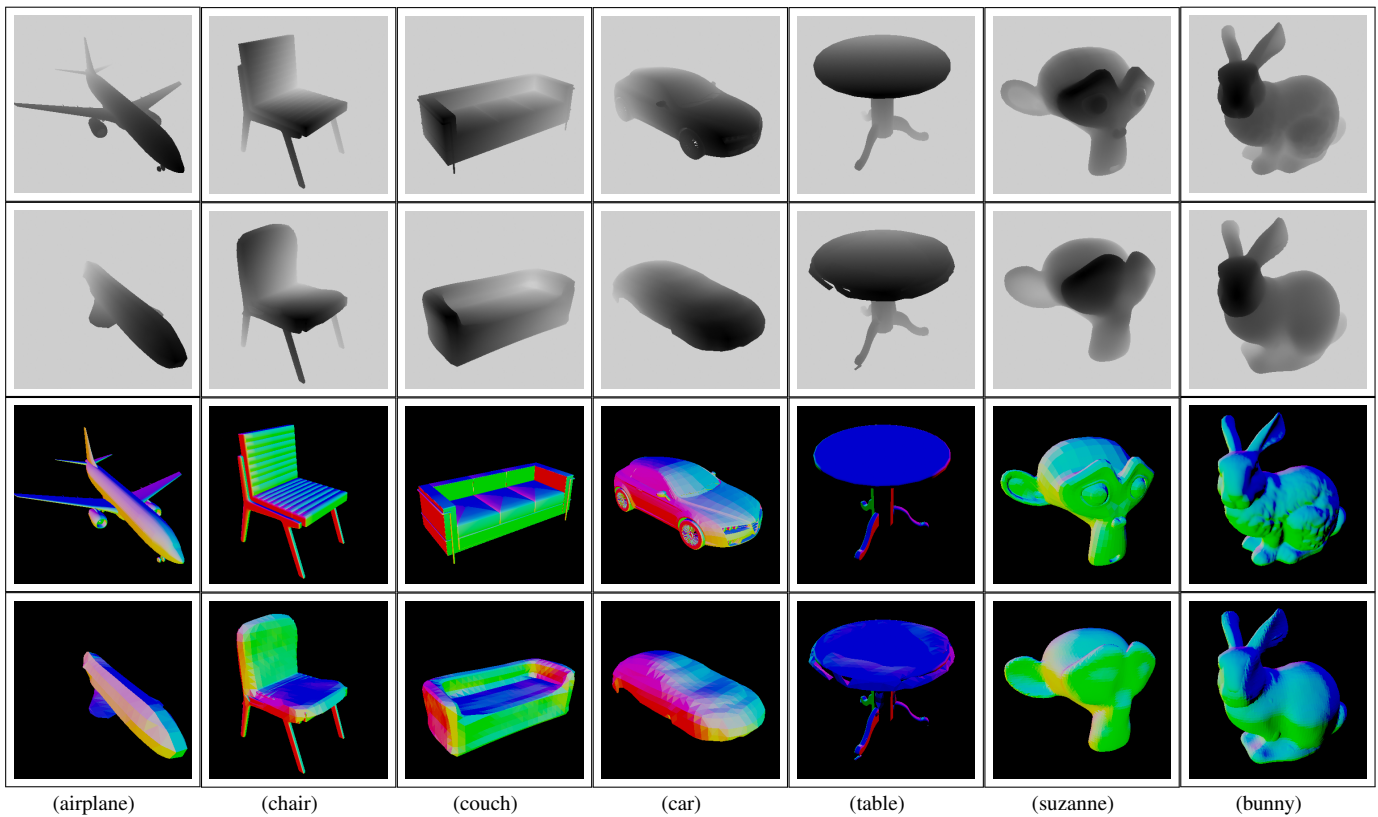
\centering
\foreach \i in {airplane,chair,couch,car,table,suzanne,bunny} {
    \begin{minipage}{.13\linewidth}\centering
    \fbox{\includegraphics[width=\linewidth]{./img/depth_normal/\i_depth_gt.png}}
    \fbox{\includegraphics[width=\linewidth]{./img/depth_normal/\i_depth_pred.png}}
    \fbox{\includegraphics[width=\linewidth]{./img/depth_normal/\i_normal_gt.png}}
    \fbox{\includegraphics[width=\linewidth]{./img/depth_normal/\i_normal_pred.png}}
    {\footnotesize (\i)} \end{minipage}
}
\caption{Comparison of the predicted in different categories from top to bottom:
ground truth field values, predicted field values, ground truth normal and predicted
normal for different categories: airplane, chair, couch, car, table, suzanne and
bunny. (Note, that this is in Z-up co-ordinate system)}
\label{fig_result}
\end{figure*}
\begin{figure*}[h]
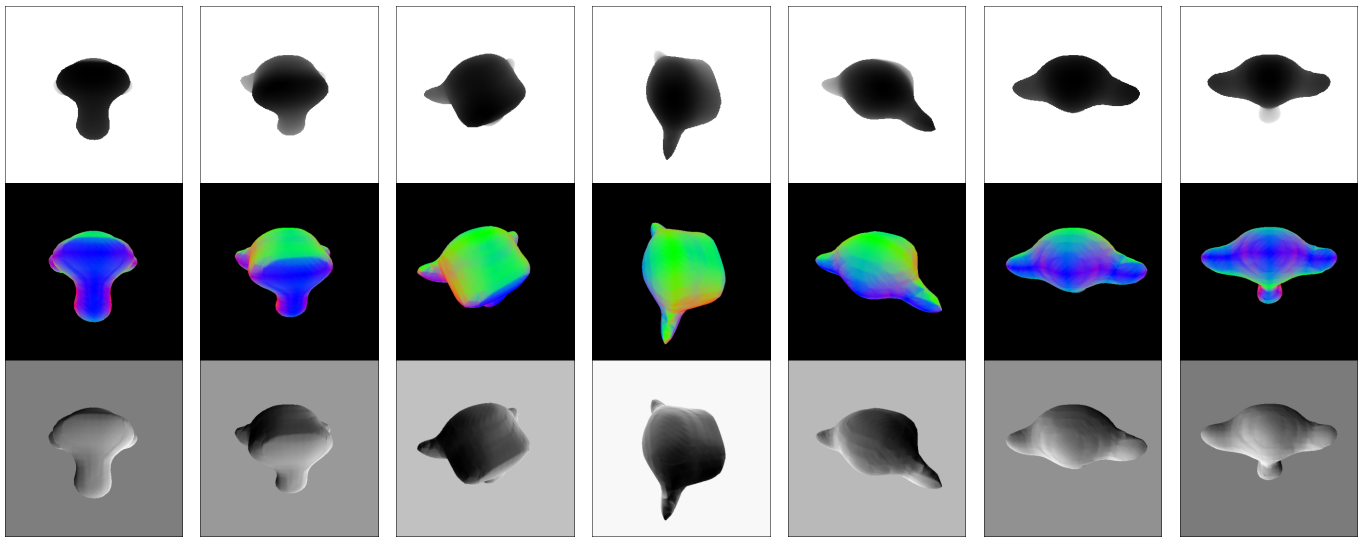
\centering
\foreach \i in {02, 06, 11, 16, 21, 26, 30} {
    \begin{minipage}{.13\linewidth}\centering
    \includegraphics[width=\linewidth]{./img/renders/depth0\i.png}
    \includegraphics[width=\linewidth]{./img/renders/normal0\i.png}
    \includegraphics[width=\linewidth]{./img/renders/rt0\i.png}
    \end{minipage}
}
\caption{The output of our renderer for $t$-values (top), normal map (middle)
and ray-traced (bottom) rendering from a sample different azimuthal and
polar angles.}
\label{fig_render_result}
\end{figure*}
\begin{figure}[h]\centering
    \includegraphics[width=\linewidth]{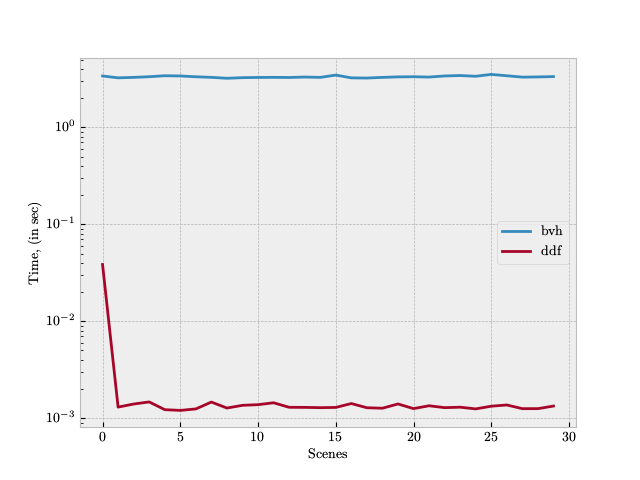}
\caption{The time taken to render images, by DDF and by
BVH. Rendering took more time the first time around because, the data
need to be copied to the GPU and cached. After that, the neural network
is cached on the GPU, the subsequent render time decreases. Note that
the y-axes is in log-scale.}
\label{fig_render_time}
\end{figure}
\begin{figure}
    \includegraphics[width=\linewidth]{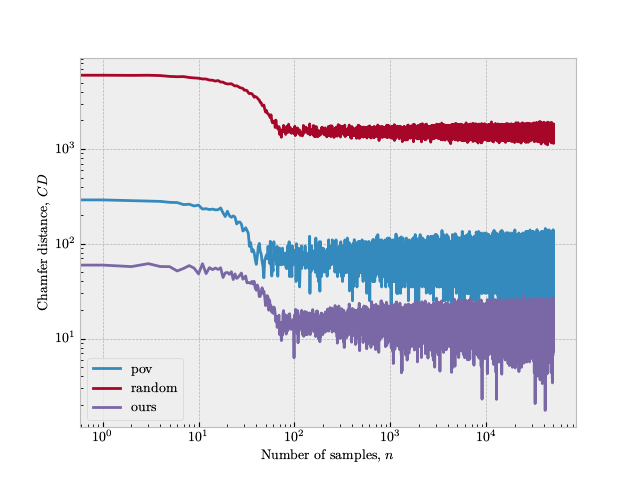}
    \caption{The chamfer distance $CD$ of the learned DDF field, against the
    number of samples. Note that this is a log-log plot of different sampling
    techniques: random, pov and ours.}
    \label{fig_sample_plot}
\end{figure}
\begin{table}\centering
\caption{\rm Experimental comparison of shape reconstruction of DeepSDF,
DeepDDF and ours with chamfer distance (in multiple of $10^3$) against
the ground truth (lower is better).}
\begin{tabular}{c|cccc}
    Object   & DIST\cite{liu2020dist} & DeepSDF\cite{park2019deepsdf} & DeepDDF\cite{yoshitake23} & Ours \\
    \hline\hline
    Chair    & 0.560 & {\bf 0.512} & 0.533 & 0.519\\
    Car      & 0.521 & 0.475 & {\bf 0.430} & 0.681\\
    Table    & 0.738 & 0.655 & {\bf 0.611} & 0.701\\
    Couch    & 0.731 & \bf 0.692 & 0.693 & {\bf 0.665}\\
    Airplane & 0.831 & 0.792 & {\bf 0.735} & 0.951\\
    Bunny    & -- & -- & --          & {\bf 0.589} \\
    Suzanne  & -- & -- & --          & {\bf 0.603} \\\hline
\end{tabular}
\label{table_comparision}
\end{table}
\subsection{Ablation study}
We evaluate our methodology against different hyperparameter by the following
metrics between the reconstructed mesh and the ground truth values: (a) chamfer
distance,
\begin{equation}
\begin{split}
\text{CD}(P_1, P_2) = &\dfrac{w_1}{|P_1|}\sum\limits_{p_{1i} \in
P_1}\min\limits_{p_{2j} \in P_2}(||p_{1i} - p_{2j}||_2^2) +\\
&\dfrac{w_2}{|P_2|}\sum\limits_{p_{2j} \in P_2}\min\limits_{p_{1i} \in
P_1}(||p_{2j} - p_{1i}||_2^2)
\end{split}
\end{equation}
(b) the f-score, and (c) the sided distance,
\begin{align}
\text{SD}(p_{1i}, P_2)=\min\limits_{p_{2j}\in{P_2}}(||p_{1i} - p_{2j}||_2^2)
\end{align}
\begin{figure}\centering
\foreach \i in {0, 45, ..., 315} {
    \begin{minipage}{.22\linewidth}\centering
    \fbox{\includegraphics[width=\linewidth]{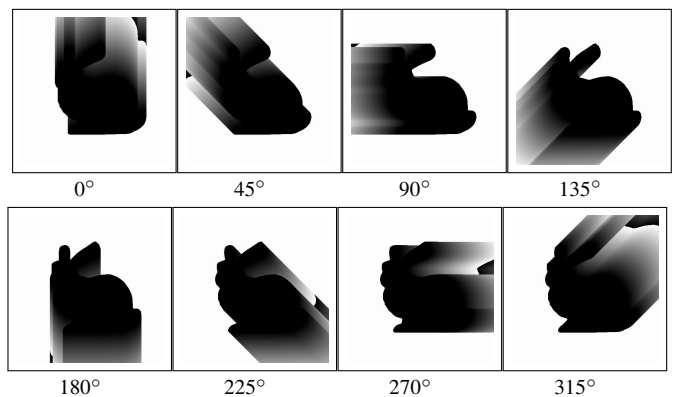}}\\
    {\footnotesize \i$^\circ$}
    \end{minipage}\vspace{1pt}
}
\caption{Shows the field values for a given cross-section of bunny rabbit for
for different polar angles. The grayscale intensity denotes the $t$-value with
black as zero.}
\label{fig_dataset}
\end{figure}
All of these metrics are computed on sets of points (or point cloud).
The DDF was converted into UDF (Eq. \ref{eq_udf}), which in turn was
converted into a triangle mesh using the marching cubes algorithm from
(Sci-kit image library\cite{van2014scikit}) and computed using the
Kaolin library\cite{kaolin}. { Table. \ref{table_hyperparam} show the
hyperparameters $s_\text{fc}$, $s_\text{dr},$ and $s_\text{p}$ against
the above metrics where we observer that increasing the samples per
face $s_\text{fc}$ has less effect, than increasing the $s_\text{p}$
and lesser still for $s_\text{dr}$.}

\begin{table}\centering
\caption{\rm Hyperparameter sensitivity: $s_\text{fc}$, $s_\text{dr},$
and $s_\text{p}$ against CD (in multiple $10^3$, lower is better).}
\begin{tabular}{c|ccc}
    n & $s_\text{fc}$ & $s_\text{dr}$ & $s_\text{p}$\\
    \hline\hline
    10 & 0.714 & 0.944 & 0.832\\
    20 & 0.829 & 0.723 & 0.019\\
    30 & 0.694 & 0.564 & 0.005\\\hline
\end{tabular}
\label{table_hyperparam}
\end{table}

\subsection{Results} {For code and instruction on how to reproduce
the results of this paper, one may visit the following URL,
\texttt{www.github.com/smlab-niser/23ddf}}. We evaluate the above theories
on a set of following triangle mesh from the ShapeNet\cite{shapenet}
dataset: airplane, car, chair, couch and table. We have also tested
the result on two standard test meshes: Standford bunny and Blender
Suzanne monkey. All the training and experimentation was done on a
Ryzen Threadripper 3970X CPU with RTX 3090 GPU. Building the dataset
took 5 hours for each mesh, while the training the neural network took
20 hours for 10,000 iteration.  The reconstructed mesh is given in
Fig. \ref{fig_result}. We have compared the chamfer distance against the
similar work in Table. \ref{table_comparision}. Although, our methodology
performs closer to similar works, it performs as good with much less
number of iterations and while also allowing for faster rendering.

{ Fig. \ref{fig_sample_plot} shows the chamfer distance,
distance between the ground truth ShapeNet and the reconstructed mesh
for different sampling techniques. For random, points and directions were
picked from the bounding box at random, a ray was shot in the direction
and the $t$-value was stored. For pov (point-of-view) sampling, we set
up cameras pointed at the mesh center around the mesh, by uniformly
spacing 10 cameras around azimuthal and 5 cameras around polar angles,
for a total of 50 cameras. For every pixel in the camera film, we shot
rays and calculated the $t$ values. We observe that out sampling works
better than the other techniques.}

{ We demonstrate ray-tracing with our renderer for
depth ($t$-value) and normal values using DDF and without
DDF that uses BVH to search for hit point and compare the
time taken Fig. (\ref{fig_render_time}) and sample renders
Fig. (\ref{fig_render_result}) for bunny. We observe that our renderer
works very much faster with DDF than without.}

\section{Conclusion} Due to pervasiveness of neural networks as
universal function estimators and learning algorithms estimate signed
distance functions, unit gradient functions, radiance fields are getting
increasing popular for representing objects. In this paper, we have
tested the use of directed distance functions as a way to represent
objects. We have tested DDF against SDFs and conclude that with similar
network training cost with respect to both computation time and parameter
complexity, the DDF produce shapes as accurate as SDFs. But for render
times, DDFs are very fast when compared to SDFs or other
explicit representations like triangle meshes.  While we have explored
only object representations with DDF, we plan to use DDFs to represent
entire scenes so that we can do faster path traced rendering.

In games, movies and virtual/augmented reality applications, where the
scene is mostly static with a few moving parts, such as a player moving
through a room in a game, directed distance fields can be used as a {\it
baking} technique to speed up the rendering of the scene. The irradiance
need to be computed only for the rays hitting dynamic objects. For scenes
like this, the rendering time can reduced considerably by using DDF. This
work focuses on representing single objects and in the future, we want to
test feasibility of DDFs as representation of entire scenes. Precomputed
radiance fields like NeRF\cite{2021nerf} also store field values of a
scene, however they require volumetric ray marching, while DDFs can do
accurate surface rendering.

\bibliographystyle{IEEEtran.bst}
\bibliography{ref}

\end{document}